\documentclass[%
 aip,
 apl,%
 amsmath,amssymb,
 reprint,%
]{revtex4-1}

\usepackage{graphicx}
\usepackage{dcolumn}
\usepackage{bm}

\begin{document}

\preprint{AIP/123-QED}

\title{Enhanced photon blockade in an optomechanical system with parametric amplification}

\author{Dong-Yang Wang}
\affiliation{Department of Physics, Harbin Institute of Technology, Harbin, Heilongjiang 150001, China}
\author{Cheng-Hua Bai}
\affiliation{Department of Physics, Harbin Institute of Technology, Harbin, Heilongjiang 150001, China}
\author{Xue Han}
\affiliation{Department of Physics, Harbin Institute of Technology, Harbin, Heilongjiang 150001, China}
\author{Shutian Liu}
\email{stliu@hit.edu.cn}
\affiliation{Department of Physics, Harbin Institute of Technology, Harbin, Heilongjiang 150001, China}
\author{Shou Zhang}
\email{szhang@ybu.edu.cn}
\affiliation{Department of Physics, College of Science, Yanbian University, Yanji, Jilin 133002, China}
\author{Hong-Fu Wang}
\email{hfwang@ybu.edu.cn}
\affiliation{Department of Physics, College of Science, Yanbian University, Yanji, Jilin 133002, China}

\begin{abstract}
We propose a scheme to enhance the single- and two-photon blockade effect significantly in a standard optomechanical system (OMS) via optical parametric amplification (OPA). The scheme does not rely on the strong single-photon optomechanical coupling and can eliminate the disadvantages of suppressing multi-photon excitation incompletely. Through analyzing the single-photon blockade (1PB) mechanism and optimizing the system parameters, we obtain a perfect 1PB with a high occupancy probability of single-photon excitation, which means that a high quality and efficient single-photon source can be generated. Moreover, we find that not only the two-photon blockade (2PB) effect is significantly enhanced but also the region of 2PB occurring is widened when the OPA exists, where we also derive the optimal parameter condition to maximize the two-photon emission and the higher photon excitations are intensely suppressed at the same time.
\end{abstract}

\keywords{single- and two-photon blockade, optomechanics, parametric amplification}
\maketitle
Optomechanics~\cite{RevModPhys.86.1391,APL.107.091116,SR.6.38559,APL.107.191110}, studying the various effects of quantum mechanics on the macroscopic scale, has made great progress over the past decade. For example, the macroscopic mechanical oscillator cooling, squeezing, entanglement, etc. have been widely investigated and verified experimentally~\cite{Nature.478.89,Science.349.952}. In addition, the reaction effect of mechanical motion on the optical field has also attracted special attention. Many interesting topics have been reported one after the other, such as the optomechanically induced transparency~\cite{PhysRevA.81.041803,Science.330.1520} and PB~\cite{PhysRevLett.107.063601,PhysRevLett.107.063602,PhysRevA.88.023853,JPB.46.035502,OE.27.27649,PhysRevA.99.043818}. Among them, the PB is a nonclassical anti-bunching effect and satisfies the sub-Poissonian light statistics, which can be used to generate the single-photon source and is particularly important for some fundamental studies in quantum optics and quantum information processing fields. Moreover, the study of multi-photon blockade~\cite{PhysRevA.87.023809,PhysRevLett.118.133604,PhysRevA.100.053857} is reported in recent years. As the name implies, the $n$PB means that the generation of the $n$-th photon will block the emergence of the $(n+1)$-th photon, which results in the $n$-th order super-Poissonian or Poissonian photon statistics and $(n+1)$-th order sub-Poissonian photon statistics.

To achieve PB, researchers have proposed two different physical mechanisms, which respectively rely on the anharmonic eigenenergy spectrum~\cite{PhysRevA.46.R6801,PhysRevLett.79.1467,PhysRevA.90.023849} and the destructive quantum interference between excitation paths~\cite{PhysRevLett.104.183601,PhysRevA.96.053827,PhysRevLett.122.243602,OL.43.5086} and are so-called conventional and unconventional PB. Specifically, the anharmonic eigenenergy spectrum in the convention PB mechanism usually comes from kinds of nonlinearities. So the achievement of strong PB requires a very large nonlinear strength, which is the major obstacles for experimentally achieving the perfect PB. On the other hand, although the two-photon excitation state can be completely suppressed when the destructive quantum interference is satisfied, the multi-photon excitations cannot be eliminated simultaneously in the unconventional PB mechanism, which result in a reduction of the PB effect~\cite{PhysRevA.96.053810}. In usual OMSs, the 1PB based on the anharmonic eigenenergy spectrum is first proposed and studied in Refs.~\cite{PhysRevLett.107.063601,PhysRevLett.107.063602}, where the anharmonicity originates from the nonlinear optomechanical coupling. Therefore, achievement of strong PB requires a large enough single-photon optomechanical coupling, which is still a big challenge in experiments. The unconventional PB in OMS has also been studied via introducing an auxiliary cavity~\cite{JPB.46.035502,PhysRevA.87.013839}. However, the required coupling between photons is too large to satisfy the condition of complete destructive quantum interference.

Here we investigate the highly significant enhancement effect of single- and two-photon blockade in an usual OMS trapping an OPA. Through calculating the second- and third-order correlation functions and those occupancy probabilities of different photon excitation number, we analytically optimize the parameter condition of respective PB cases (1PB and 2PB). We show that, without requiring the strong optomechanical coupling condition as in the usual OMS, a perfect 1PB can be achieved and the multi-photon excitations are also suppressed completely. Meanwhile, the single-photon occupancy probability is maximized via analyzing the resonance condition, which indicates that the efficiency of single-photon emission is the highest at this time. Furthermore, the enhanced 2PB is also discussed in detail with the optimal parameter condition. And we find that the existence of OPA not only enhances the 2PB effect, but also widens the region of 2PB occurring.

\begin{figure}
	\centering
	\includegraphics[width=0.9\linewidth]{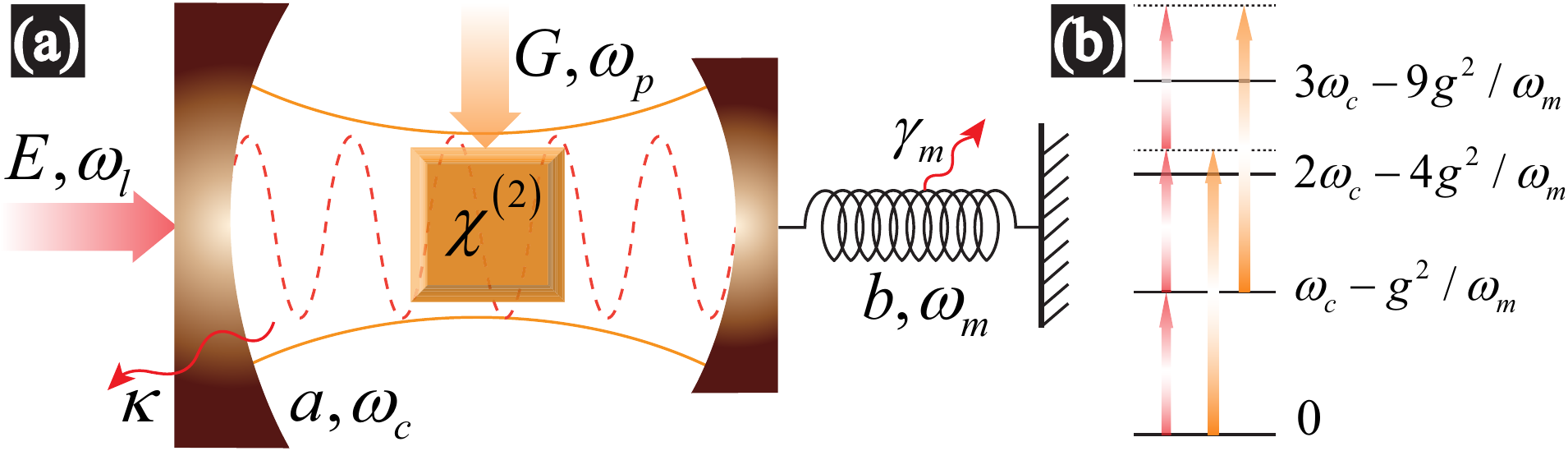}
	\caption{(a) Schematic diagram of an OMS with a $\chi^{2}$-type nonlinear medium in the cavity. The cavity is pumped by a classical laser and the nonlinear medium is driven by a pumping field. (b) The anharmonic eigenenergy spectrum with different photon excitation number.}\label{fig:optomechanicsOPA}
\end{figure}

We consider an OMS trapping a $\chi^{2}$-type nonlinear medium~\cite{NonlinearOptics,APL.112.241105,OL.43.2050,PhysRevA.93.043844}, as depicted in Fig.~\ref{fig:optomechanicsOPA}, where the cavity is pumped by a classical laser with frequency $\omega_{l}$, amplitude $E$, and phase $\phi_{l}$. The nonlinear medium is a degenerate OPA, which is driven by a pumping field with frequency $\omega_{p}=2\omega_{l}$, nonlinear gain rate $G$, and phase $\phi_{p}$. The optical mode interacts with the mechanical motion via radiation pressure~\cite{APL.104.061101,PhysRevA.98.023816}. In the rotating frame, the system Hamiltonian is written as ($\hbar=1$)
\begin{eqnarray}
	H&=&\Delta_{c}a^{\dagger}a+\omega_{m}b^{\dagger}b-ga^{\dagger}a(b^{\dagger}+b)+Ge^{i\theta}a^{\dagger 2}\cr\cr
	&&+Ea^{\dagger}+\mathrm{H.c.},
	\label{e01}
\end{eqnarray}
where $a$ ($b$) is the annihilation operator for the optical (mechanical) mode. The first two terms are the free Hamiltonian, $\Delta_{c}=\omega_{c}-\omega_{l}$ is the cavity pump field detuning, and $\omega_{m}$ is the mechanical frequency. The third term describes the optomechanical interaction~\cite{OL43.6053,PRJ.7.1229} with the single-photon coupling strength $g$. The last two terms represent the external driving interactions, and $\theta=\phi_{p}-2\phi_{l}$ is the relative phase between classical driving fields. Transformed into the mechanical displacement representation defined by $V=\exp[g/\omega_{m}a^{\dagger}a(b^{\dagger}-b)]$, the optomechanical interaction can be changed to the Kerr-like nonlinear of optical cavity $g^{2}/\omega_{m}(a^{\dagger}a)^{2}$. Utilizing the fact $g\ll\omega_{m}$ in most actual systems, we can regard that the mechanical oscillator decouples with the optical cavity at this time~\cite{PhysRevA.99.043818}. Therefore, when we are only interested in the optical properties, we can ignore the mechanical parts and the Hamiltonian of optical component is rewritten as $H^{\prime}=\Delta_{c}a^{\dagger}a-g^{2}/\omega_{m}(a^{\dagger}a)^{2}+Ge^{i\theta}a^{\dagger 2}+Ea^{\dagger}+\mathrm{H.c.}$.
Clearly, the free energy spectrum of cavity is $n\Delta_{c}-n^{2}g^{2}/\omega_{m}$ due to the optomechanical coupling shifting the eigenenergy spectrum, see Fig.~\ref{fig:optomechanicsOPA}(b).

Utilizing the reduced Hamiltonian, the dynamical evolution of the photon state is in the charge of the non-Hermitian Schr\"odinger equation $i\partial|\psi(t)\rangle/\partial t=(H^{\prime}-i\kappa/2a^{\dagger}a)|\psi(t)\rangle$, where $|\psi(t)\rangle=\sum_{n}C_{n}(t)|n\rangle$ is the time-dependent photon state, $C_{n}(t)$ represents the probability amplitude of $n$ ($n\geqslant0$) photons, and $\kappa$ is the cavity decay. Then we obtain a set of linear differential equations
\begin{eqnarray}
	&&i\dot{C}_{n}=-nM_{n}C_{n}/2+E\left(\sqrt{n}C_{n-1}+\sqrt{n+1}C_{n+1}\right)\cr\cr
	&&+G\left[\sqrt{n(n-1)}e^{i\theta}C_{n-2}+\sqrt{(n+1)(n+2)}e^{-i\theta}C_{n+2}\right],\cr\cr
	&&\label{e03}
\end{eqnarray}
where $M_{n}=2ng^{2}/\omega_{m}-2\Delta_{c}+i\kappa$. Under the weak driving condition $\{E,~G\}\ll\kappa$, we can truncate the space of photon state by a low-excitation number, e.g., $n\leqslant3$. The steady-state solution of probability amplitudes thus is approximatively given by
\begin{eqnarray}
	C_{1}&\simeq&2E/M_{1},~~~~C_{2}\simeq\sqrt{2}\left(2E^{2}+Ge^{i\theta}M_{1}\right)/M_{1}M_{2},\cr\cr
	C_{3}&\simeq&\frac{2\sqrt{2}E\left[2E^{2}+Ge^{i\theta}\left(4Ge^{i\theta}-2g^{2}/\omega_{m}+3M_{2}\right)\right]}{\sqrt{3}M_{1}M_{2}M_{3}}.
	\label{e04}
\end{eqnarray}
The 1PB is characterized by the correlation function
\begin{eqnarray}
	g^{(2)}(\tau)=\frac{\langle a^{\dagger}(t)a^{\dagger}(t+\tau)a(t+\tau)a(t)\rangle}{\langle a^{\dagger}(t)a(t)\rangle\langle a^{\dagger}(t+\tau)a(t+\tau)\rangle},
	\label{e05}
\end{eqnarray}
which is the probability of detecting two photons spaced with a finite-time delay $\tau$. So the mathematical representation of 1PB is $g^{(2)}(0)<1$. Substituting the steady-state solution into the equal-time correlation function $g^{(2)}(0)$, we can obtain the analytical result $g^{(2)}(0)\simeq\left(2|C_{2}|^{2}+6|C_{3}|^{2}\right)/|C_{1}|^{4}$.
Here, we reserve the influence of higher excitation ($n=3$) on 1PB to test whether the multi-photon excitation is suppressed completely.

On the other hand, the exact numerical simulation can be carried out by utilizing the quantum master equation with the initial Hamiltonian $H$
\begin{eqnarray}
	\dot{\rho}&=&-i\left[H,\rho\right]+\kappa\mathcal{L}[a]\rho+\gamma_{m}\mathcal{L}[b]\rho,
	\label{e07}
\end{eqnarray}
where $\rho$ is the system density matrix, $\gamma_{m}$ represents the mechanical damping rate, and $\mathcal{L}[o]\rho$ is the standard Lindblad operator for the arbitrary system operator $o$. When the system reaches its steady state $\rho_{s}$, the exact numerical result is given by $g^{(2)}(0)=\mathrm{Tr}[a^{\dagger}a^{\dagger}aa\rho_{s}]/\mathrm{Tr}[a^{\dagger}a\rho_{s}]^{2}$.

\begin{figure}
	\centering
	\includegraphics[width=0.4\linewidth]{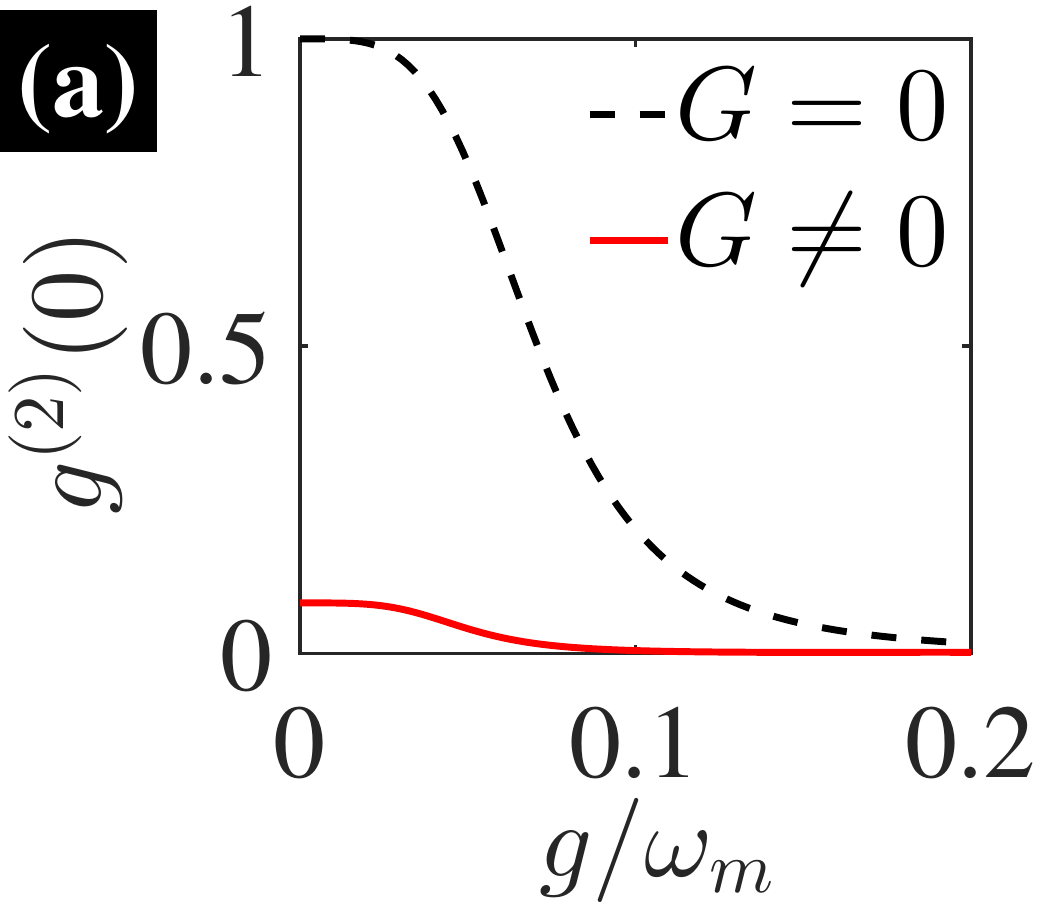}
	\includegraphics[width=0.4\linewidth]{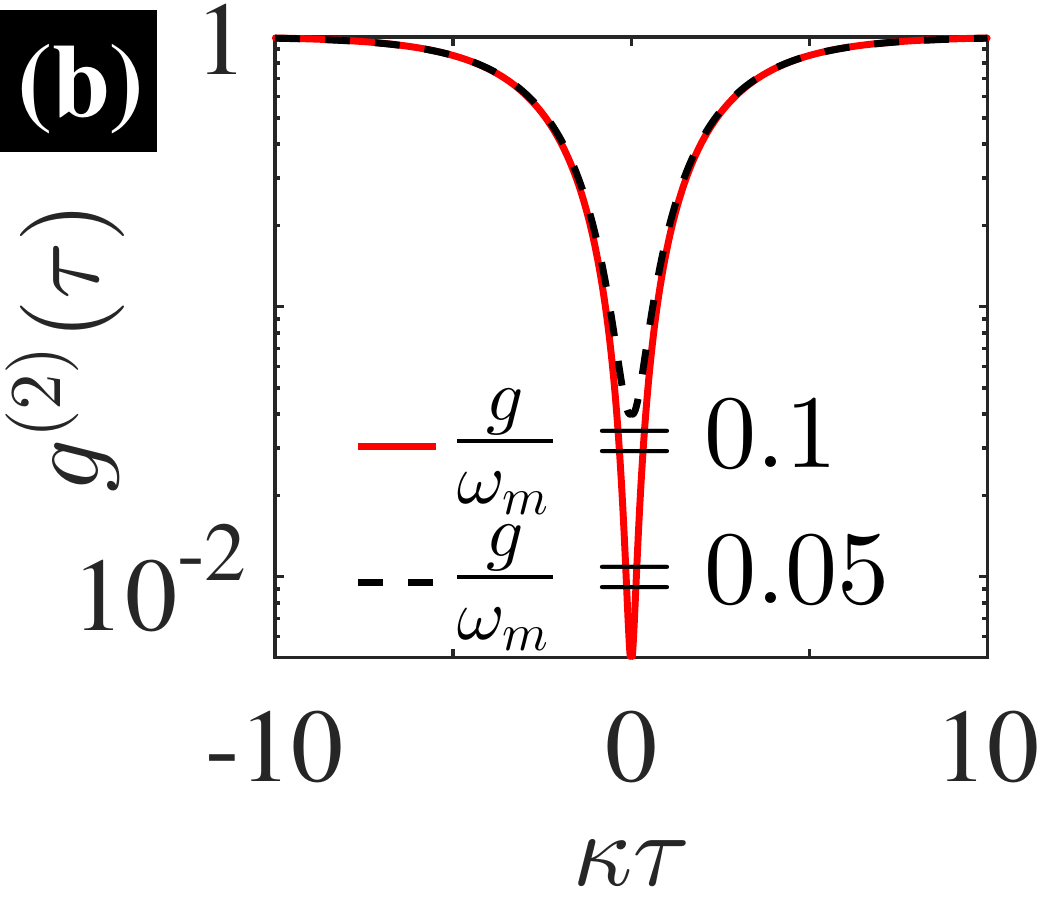}
	\includegraphics[width=0.4\linewidth]{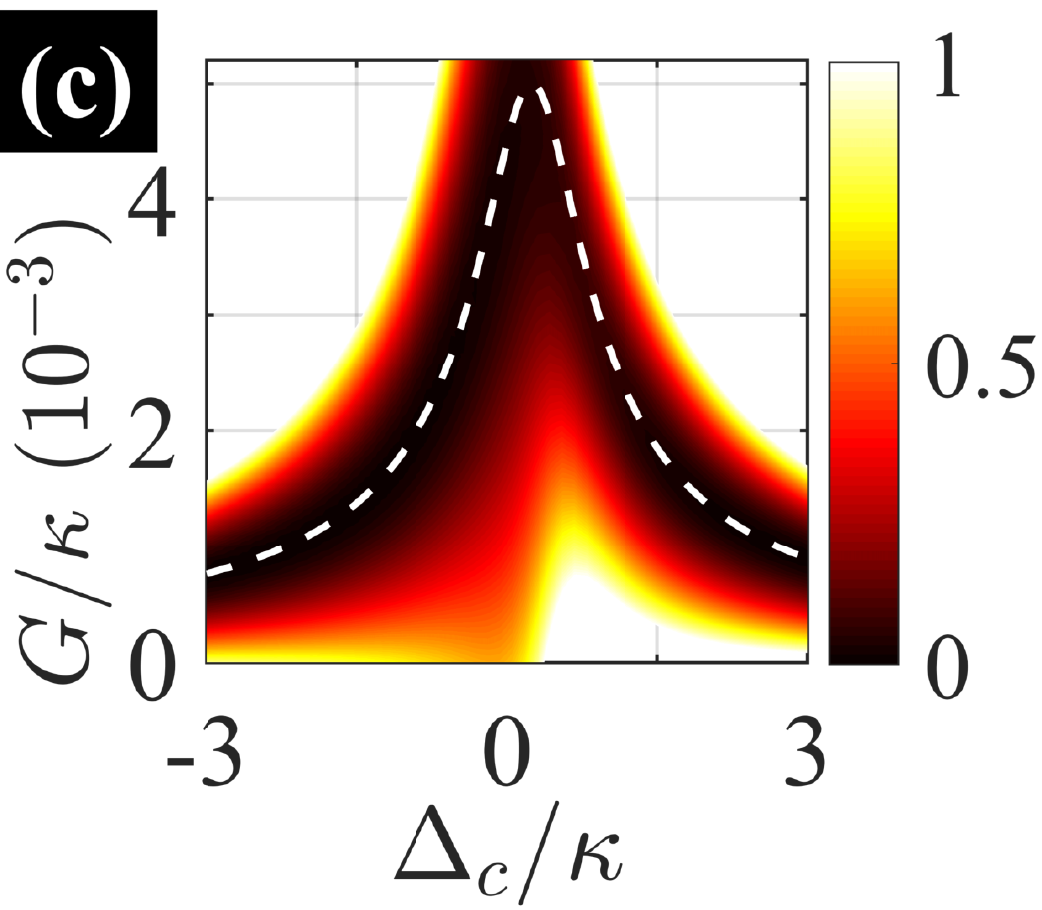}
	\includegraphics[width=0.4\linewidth]{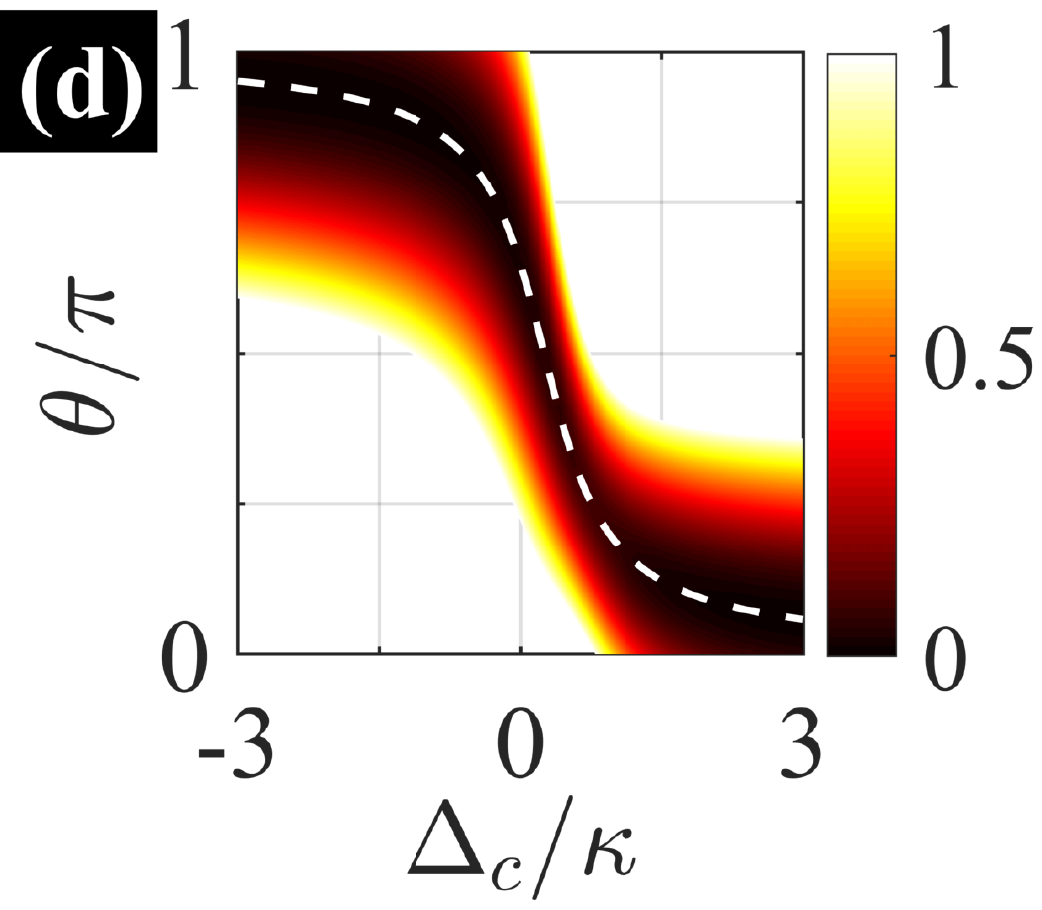}
	\caption{The effect of system parameters on 1PB. (a) The enhancement of 1PB via OPA. (b) The delayed second-order correlation function with different optomechanical couplings. (c) and (d) represent the equal-time second-order correlation function versus different system parameters. The dashed white lines are the optimal conditions $G_{\mathrm{opt}}=|-2E^{2}/M_{1}|$ and $\theta_{\mathrm{opt}}=\mathrm{Arg}[-2E^{2}/M_{1}]$, respectively.}\label{fig:1PB}
\end{figure}

Next, we concretely discuss the enhanced 1PB effect and photon statistics property based on the above analytical and numerical calculations. Before our discussion, we analyze the optimal condition of 1PB occurring, i.e., $Ge^{i\theta}=-2E^{2}/M_{1}$. Under the optimal condition, we first verify the enhancement effect of 1PB via OPA, as shown in Fig.~\ref{fig:1PB}(a). We can see that, owning to the existence of OPA ($G\neq0$), the 1PB effect is significantly enhanced even in the weak optomechanical coupling region, which breaks the restriction of strong coupling as in optomechanical systems~\cite{PhysRevLett.107.063601,PhysRevLett.107.063602}. However, we also notice the enhanced 1PB is imperfect when the optomechanical coupling is too small. That is because the small coupling cannot cause a sufficient anharmonic eigenenergy spectrum, resulting in the incomplete suppression of multi-photon excitation. As the coupling increases, the 1PB will be gradually perfect $g^{(2)}(0)=0$. Furthermore, the delayed second-order correlation function is plotted with different optomechanical couplings, see Fig.~\ref{fig:1PB}(b). We find that $g^{(2)}(\tau)$ is always larger than $g^{(2)}(0)$ and reaches 1 when the delay time is long enough. To verify the optimal condition, we plot the correlation function versus different system parameters in Figs.~\ref{fig:1PB}(c) and \ref{fig:1PB}(d), respectively. In the above simulation, the other system parameters are chosen as $g/\omega_{m}=0.05$, $\omega_{m}/\kappa=100$, $\gamma_{m}/\omega_{m}=10^{-6}$, and $E/\kappa=0.05$ that satisfy the approximate conditions we need $\{E,~G\}\ll\kappa$ and $g\ll\omega_{m}$.

\begin{figure}
	\centering
	\includegraphics[width=0.4\linewidth]{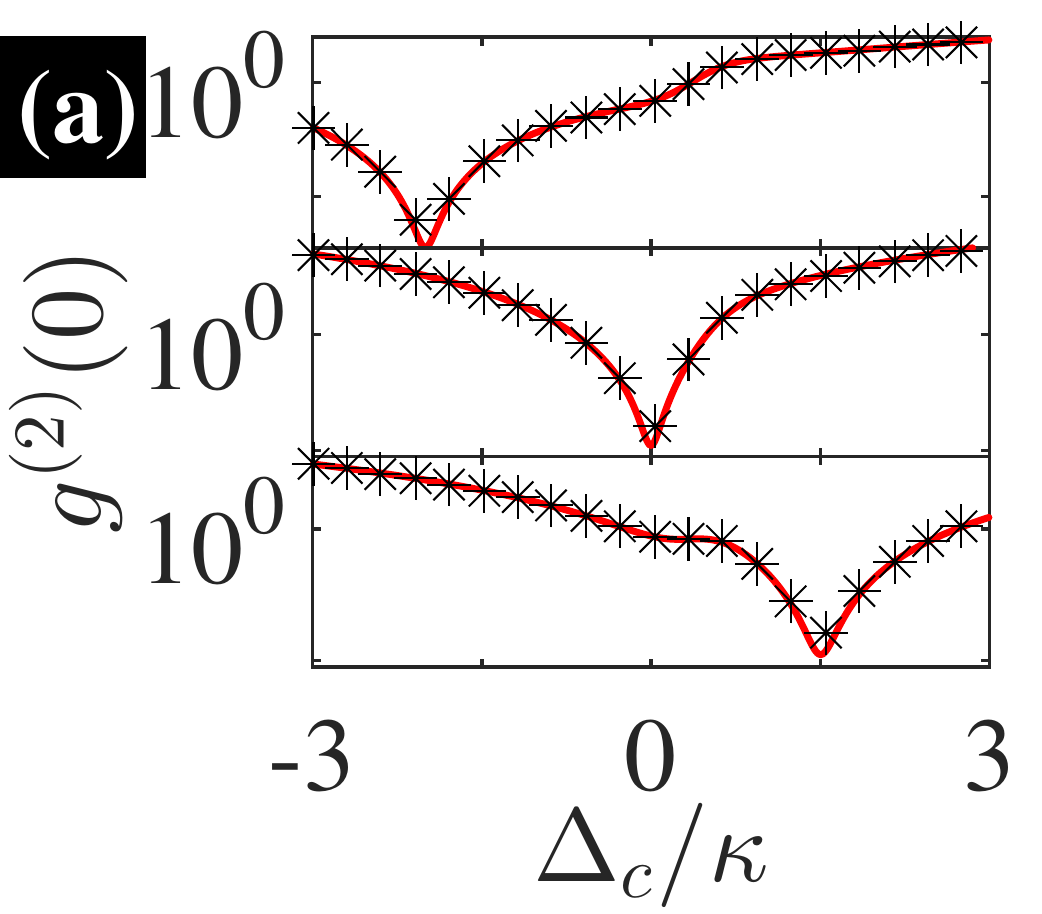}
	\includegraphics[width=0.4\linewidth]{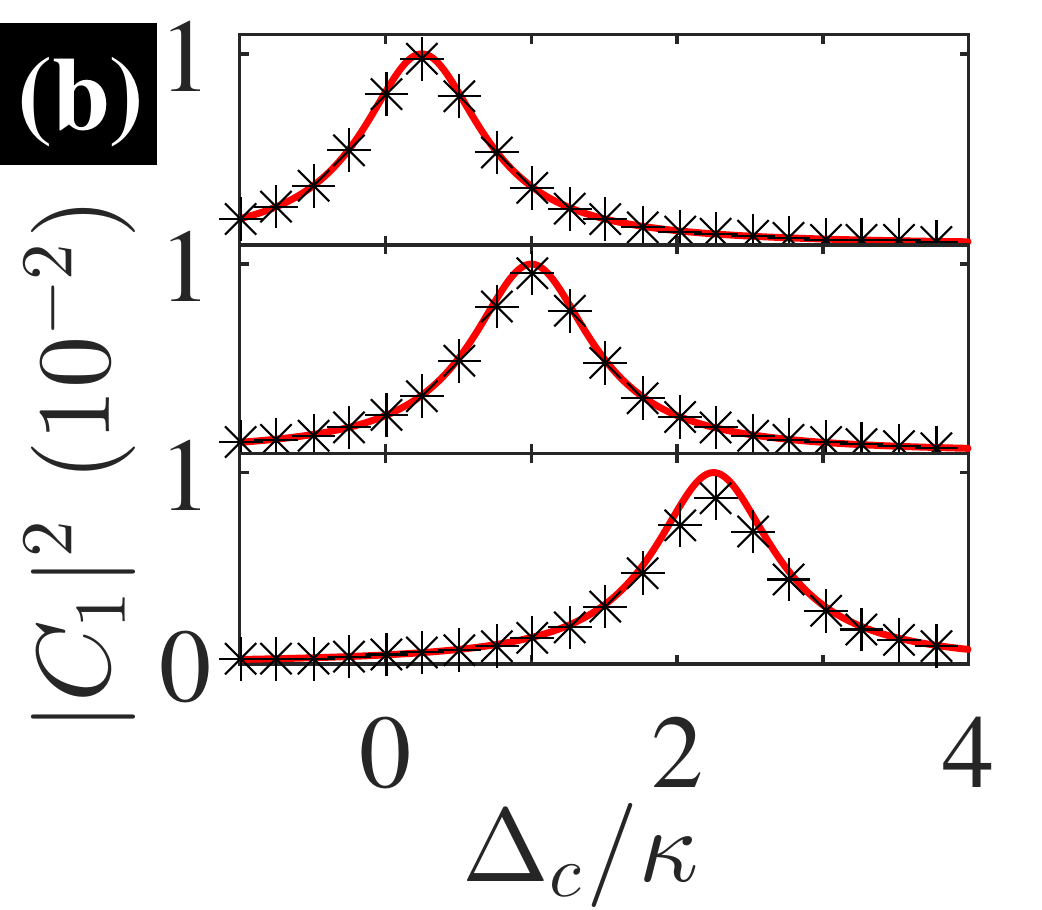}
	\caption{Analytical solution (red line) and numerical simulation (black asterisk). (a) $g^{(2)}(0)$ versus the cavity pump field detuning with different pairs of optimal parameters, which respectively result in the perfect 1PB occurring at $\Delta_{c}/\kappa=\{-2,~0,~1.5\}$. (b) The single-photon occupancy probability versus the cavity pump field detuning with different optomechanical coupling strengths $g/\omega_{m}=\{0.05,~0.1,~0.15\}$.}\label{fig:1PBopt}
\end{figure}

According to the above analyses, the location of perfect 1PB can be modulated via changing the optimal parameters $G_{\mathrm{opt}}$ and $\theta_{\mathrm{opt}}$, as shown in Figs.~\ref{fig:1PB}(c) and \ref{fig:1PB}(d). Specially, the perfect 1PB occurring at different $\Delta_{c}$ is shown in Fig.~\ref{fig:1PBopt}(a) by analytical and numerical calculations, respectively. Here, the optimal parameters have been chosen according to the above optimal condition. It seems that the perfect 1PB at any location is possible. However, we cannot ignore the fact that the perfect 1PB also requires a high single-photon occupancy probability to achieve the high efficient single-photon source. Whether the perfect 1PB at any location all has the high occupancy probability needs to be further verified. We calculate the single-photon occupancy probability with different optomechanical couplings when the perfect 1PB occurs and show the results in Fig.~\ref{fig:1PBopt}(b). We find that the distribution of occupancy probability satisfies the Lorentz linearity and its peak is located at $\Delta_{c}/\kappa=g^{2}/\omega_{m}$, which corresponds to the single-excitation resonant condition, see Fig.~\ref{fig:optomechanicsOPA}(b). That means the single-photon occupancy probability of perfect 1PB is not always high for any location. With the enhancement of optomechanical coupling, the location of perfect 1PB with a high single-photon occupancy probability moves to the right and the difference between analytical and numerical results arises gradually. That is because the excessively large optomechanical coupling leads to the fail of approximative analytical solution. Combined with the OPA and optomechanical coupling, the perfect 1PB is obtained and the single-photon occupancy probability is almost $10^{-2}$, which indicates that the efficiency of single-photon emission can reach $10^{4}$ per second when the decay of cavity is megahertz.

Moreover, we also explore the existence of 2PB~\cite{PhysRevA.87.023809,PhysRevLett.118.133604,PhysRevA.100.053857}, which can be judged by the criterion of $g^{(2)}(0)\geqslant1$ and $g^{(3)}(0)<1$. Here, $g^{(3)}(0)$ is the equal-time third-order correlation function defined as $g^{(3)}(0)=\langle a^{\dagger}a^{\dagger}a^{\dagger}aaa\rangle/\langle a^{\dagger}a\rangle^{3}$, which represents the probability of detecting three photons in optical cavity at the same time. According to the analytical derivation, the optimal parameter condition of $g^{(3)}(0)=0$ is given by
\begin{eqnarray}
	Ge^{i\theta}\simeq\begin{cases}
		K+\sqrt{K^{2}-E^{2}/2},~~~~\left(\Delta_{c}/\kappa<5g^{2}/3\omega_{m}\right)\\
		K-\sqrt{K^{2}-E^{2}/2},~~~~\left(\Delta_{c}/\kappa>5g^{2}/3\omega_{m}\right)
	\end{cases}
	\label{e08}
\end{eqnarray}
where $K=g^{2}/4\omega_{m}-3M_{2}/8$. We stress that the optimal condition is different from the previous 1PB optimal condition and it cannot ensure the occurrence of 2PB, which needs $g^{(2)}(0)\geqslant1$ is satisfied at the same time.

\begin{figure}
	\centering
	\includegraphics[width=0.4\linewidth]{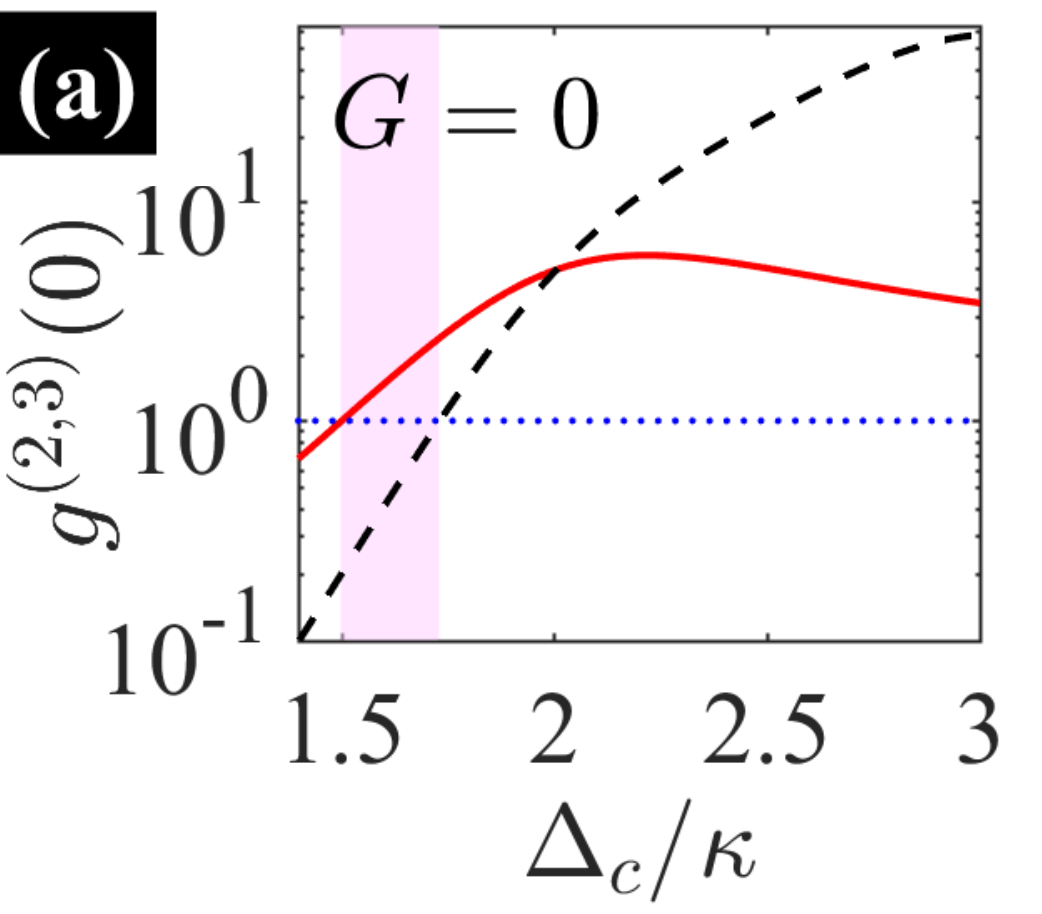}
	\includegraphics[width=0.4\linewidth]{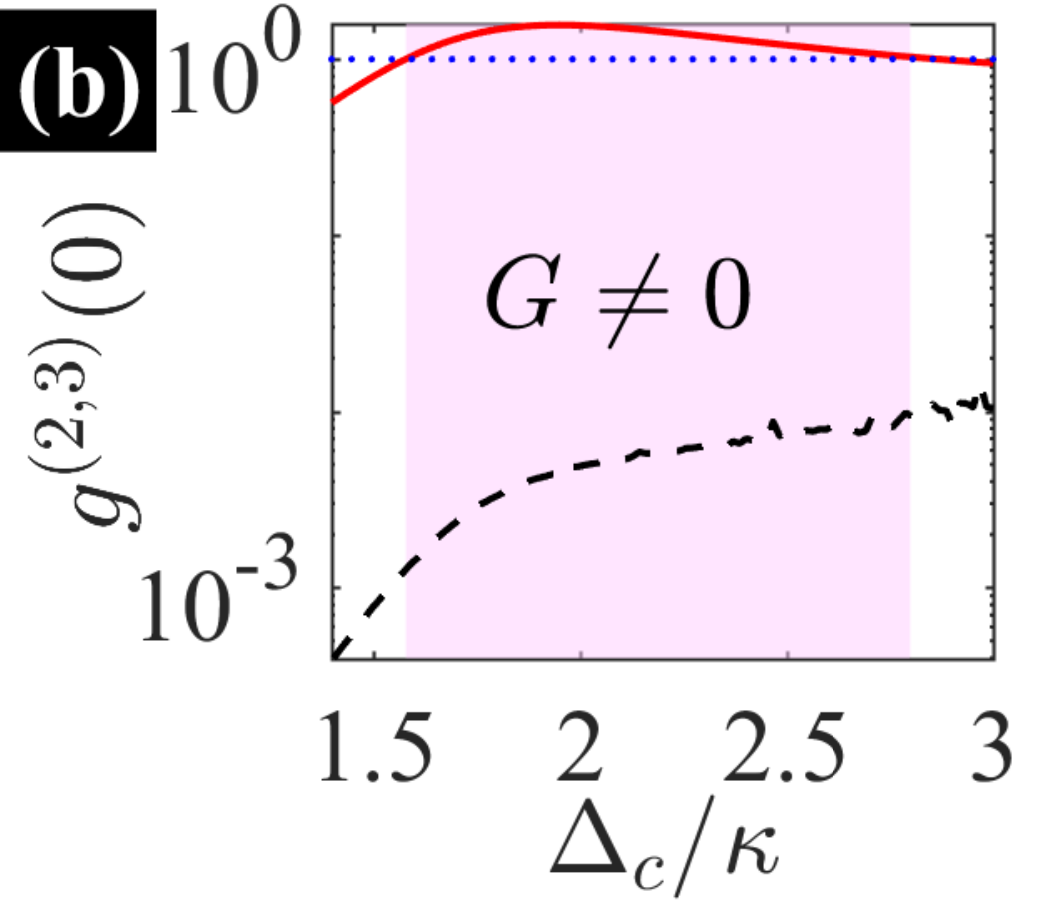}
	\includegraphics[width=0.4\linewidth]{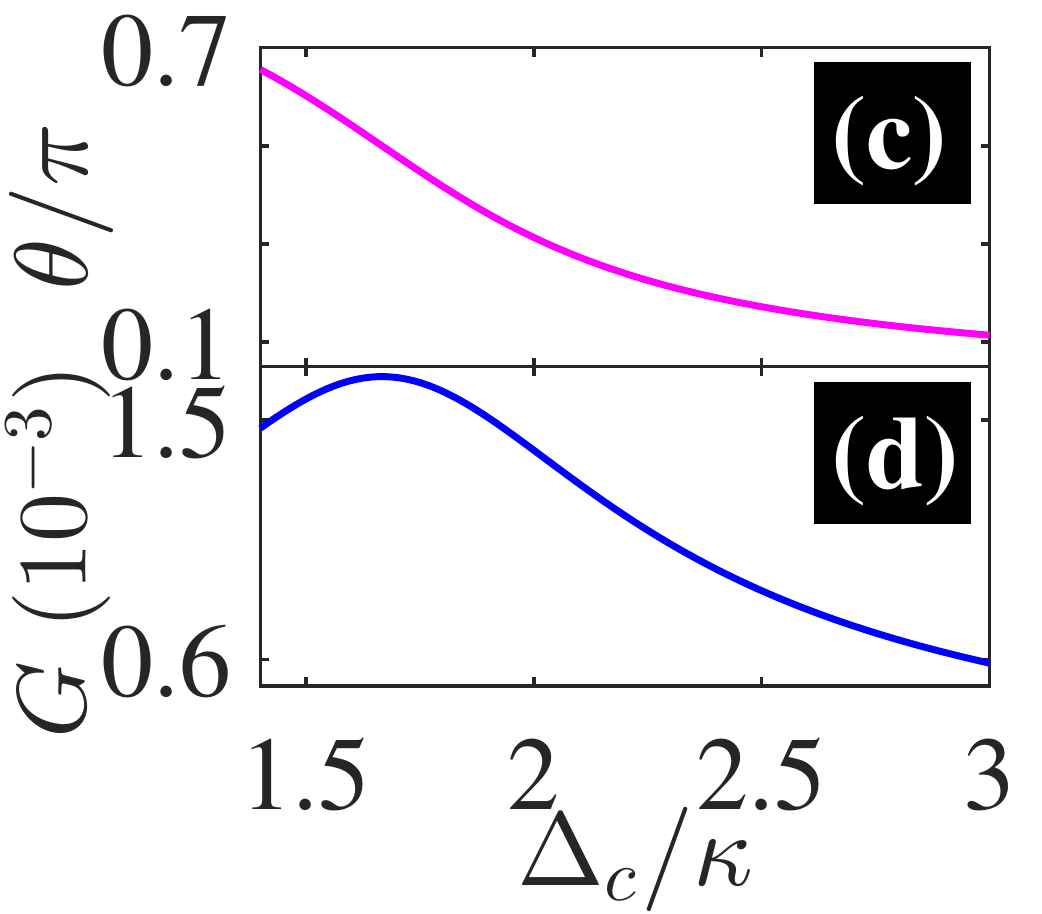}
	\includegraphics[width=0.4\linewidth]{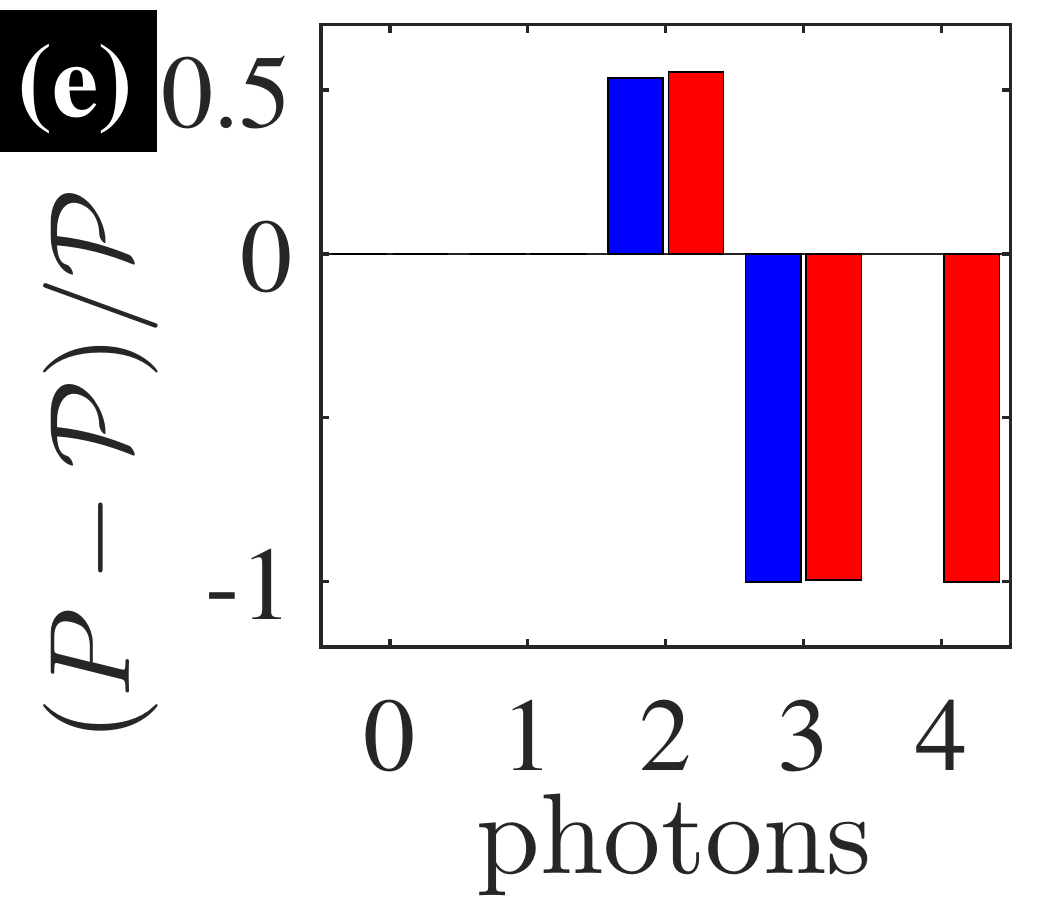}
	\caption{(a) and (b) respectively show that the region of 2PB occurring versus the cavity pump field detuning without or with OPA, where the solid red line represents the second-order correlation function, the dashed black line is the third-oder correlation function, and the pink region indicates the occurrence of 2PB, i.e., $g^{(2)}(0)\geqslant1$ and $g^{(3)}(0)<1$. (c) and (d) are the optimal $\theta$ and $G$ versus the cavity pump field detuning, as given in Eq.~(\ref{e08}). (e) The relative deviations of the analytical (blue) and numerical (red) photon distribution to the standard Poisson distribution of the same mean photon number located at $\Delta_{c}/\kappa=2$. Here, we choose $g/\omega_{m}=0.1$.}\label{fig:2PB}
\end{figure}

To verify the existence of 2PB, we respectively calculate the $g^{(2)}(0)$ and $g^{(3)}(0)$ by utilizing the derived optimal condition when the OPA exists or not, and show the results in Figs.~\ref{fig:2PB}(a) and \ref{fig:2PB}(b). We can see that the 2PB effect is significantly enhanced in the presence of OPA ($G\neq0$) and the region of 2PB occurring is increased. Under the optimal condition, the strong 2PB $\{g^{(3)}(0)\sim10^{-3},~g^{(2)}(0)\geqslant1\}$ is achieved in the vicinity of $\Delta_{c}/\kappa\simeq2$. Furthermore, we also show the optimal condition of 2PB versus the cavity pump field detuning in Figs.~\ref{fig:2PB}(c) and \ref{fig:2PB}(d). In order to further illustrate the 2PB effect, we calculate the relative deviations of the photon distribution to the standard Poisson distribution of the same mean photon number, which is defined as $(P-\mathcal{P})/\mathcal{P}$. Here, $P$ is photon-number distribution probability in our scheme, $\mathcal{P}$ represents the standard Poisson distribution probability with the same mean photon number, and the relative deviations of different photon number $n$ is given by $\mathcal{P}_{n}=e^{-\langle a^{\dagger}a\rangle}\langle a^{\dagger}a\rangle^{n}/n!$, where $\langle a^{\dagger}a\rangle$ is the mean photon number. The analytical and numerical results of the relative population located at $\Delta_{c}/\kappa=2$ are shown together in Fig.~\ref{fig:2PB}(e). We can see that the analytical result (blue) coming from Eq.~(\ref{e04}) is truncated to 3 and the numerical simulation (red) is truncated to 4 (be same to a higher truncation), where both they agree well with each other and indicate the strong 2PB occurs. Meanwhile, the higher photon excitations ($n\geqslant3$) are intensely suppressed, which also indicates the rationality of truncation we adopt. Therefore, we conclude that the two-photon emission in our model can also be significantly enhanced via the OPA while the higher excitations are effectively suppressed.

In summary, we have investigated the enhanced single- and two-photon blockade effect via the trapped OPA in a standard OMS. We find that the existences of OPA and optomechanical coupling are complementary for achieving the strong PB. Specifically, the perfect 1PB enhanced by OPA no longer requires too large single-photon optomechanical coupling, which reduces the difficulty of experimental implementation. On the other hand, the appropriate optomechanical coupling also removes the shortcoming in the incomplete suppression of multi-photon excitation. Combining optimal OPA and weak optomechanical coupling, the perfect 1PB can be effectively achieved. Meanwhile, when the perfect 1PB occurs, we also discuss how to maximize the single-photon occupancy probability to improve the efficiency of single-photon emission. For the 2PB, the OPA not only enhances the blockade effect, but also widens the region of blockade occurring. Moreover, the higher photon excitations are also intensely suppressed, which ensures the quality of 2PB. Our scheme would provide a guidance for the experimental implementation of single-photon source with a high single-photon emission efficiency and might be meaningful to investigate the multi-photon blockade.

This work was supported by the National Natural Science Foundation of China under Grant No. 61822114, No. 61575055, No. 11874132, No. 61465013, and No. 11465020, and the Project of Jilin Science and Technology Development for Leading Talent of Science and Technology Innovation in Middle and Young and Team Project under Grant No. 20160519022JH.


\begin{thebibliography}{34}%
	\makeatletter
	\providecommand \@ifxundefined [1]{%
		\@ifx{#1\undefined}
	}%
	\providecommand \@ifnum [1]{%
		\ifnum #1\expandafter \@firstoftwo
		\else \expandafter \@secondoftwo
		\fi
	}%
	\providecommand \@ifx [1]{%
		\ifx #1\expandafter \@firstoftwo
		\else \expandafter \@secondoftwo
		\fi
	}%
	\providecommand \natexlab [1]{#1}%
	\providecommand \enquote  [1]{``#1''}%
	\providecommand \bibnamefont  [1]{#1}%
	\providecommand \bibfnamefont [1]{#1}%
	\providecommand \citenamefont [1]{#1}%
	\providecommand \href@noop [0]{\@secondoftwo}%
	\providecommand \href [0]{\begingroup \@sanitize@url \@href}%
	\providecommand \@href[1]{\@@startlink{#1}\@@href}%
	\providecommand \@@href[1]{\endgroup#1\@@endlink}%
	\providecommand \@sanitize@url [0]{\catcode `\\12\catcode `\$12\catcode
		`\&12\catcode `\#12\catcode `\^12\catcode `\_12\catcode `\%12\relax}%
	\providecommand \@@startlink[1]{}%
	\providecommand \@@endlink[0]{}%
	\providecommand \url  [0]{\begingroup\@sanitize@url \@url }%
	\providecommand \@url [1]{\endgroup\@href {#1}{\urlprefix }}%
	\providecommand \urlprefix  [0]{URL }%
	\providecommand \Eprint [0]{\href }%
	\providecommand \doibase [0]{http://dx.doi.org/}%
	\providecommand \selectlanguage [0]{\@gobble}%
	\providecommand \bibinfo  [0]{\@secondoftwo}%
	\providecommand \bibfield  [0]{\@secondoftwo}%
	\providecommand \translation [1]{[#1]}%
	\providecommand \BibitemOpen [0]{}%
	\providecommand \bibitemStop [0]{}%
	\providecommand \bibitemNoStop [0]{.\EOS\space}%
	\providecommand \EOS [0]{\spacefactor3000\relax}%
	\providecommand \BibitemShut  [1]{\csname bibitem#1\endcsname}%
	\let\auto@bib@innerbib\@empty
	\bibitem [{\citenamefont {Aspelmeyer}, \citenamefont {Kippenberg},\ and\
		\citenamefont {Marquardt}(2014)}]{RevModPhys.86.1391}%
	\BibitemOpen
	\bibfield  {author} {\bibinfo {author} {\bibfnamefont {M.}~\bibnamefont
			{Aspelmeyer}}, \bibinfo {author} {\bibfnamefont {T.~J.}\ \bibnamefont
			{Kippenberg}}, \ and\ \bibinfo {author} {\bibfnamefont {F.}~\bibnamefont
			{Marquardt}},\ }\href {\doibase 10.1103/RevModPhys.86.1391} {\bibfield
		{journal} {\bibinfo  {journal} {Rev. Mod. Phys.}\ }\textbf {\bibinfo {volume}
			{86}},\ \bibinfo {pages} {1391} (\bibinfo {year} {2014})}\BibitemShut
	{NoStop}%
	\bibitem [{\citenamefont {Xiong}\ \emph {et~al.}(2015)\citenamefont {Xiong},
		\citenamefont {Si}, \citenamefont {Yang},\ and\ \citenamefont
		{Wu}}]{APL.107.091116}%
	\BibitemOpen
	\bibfield  {author} {\bibinfo {author} {\bibfnamefont {H.}~\bibnamefont
			{Xiong}}, \bibinfo {author} {\bibfnamefont {L. G.}\ \bibnamefont {Si}},
		\bibinfo {author} {\bibfnamefont {X.}~\bibnamefont {Yang}}, \ and\ \bibinfo
		{author} {\bibfnamefont {Y.}~\bibnamefont {Wu}},\ }\href {\doibase
		10.1063/1.4930166} {\bibfield  {journal} {\bibinfo  {journal} {Appl. Phys.
				Lett.}\ }\textbf {\bibinfo {volume} {107}},\ \bibinfo {pages} {091116}
		(\bibinfo {year} {2015})}\BibitemShut {NoStop}%
	\bibitem [{\citenamefont {Wang}\ \emph {et~al.}(2016)\citenamefont {Wang},
		\citenamefont {Bai}, \citenamefont {Wang}, \citenamefont {Zhu},\ and\
		\citenamefont {Zhang}}]{SR.6.38559}%
	\BibitemOpen
	\bibfield  {author} {\bibinfo {author} {\bibfnamefont {D. Y.}\ \bibnamefont
			{Wang}}, \bibinfo {author} {\bibfnamefont {C. H.}\ \bibnamefont {Bai}},
		\bibinfo {author} {\bibfnamefont {H. F.}\ \bibnamefont {Wang}}, \bibinfo
		{author} {\bibfnamefont {A. D.}\ \bibnamefont {Zhu}}, \ and\ \bibinfo
		{author} {\bibfnamefont {S.}~\bibnamefont {Zhang}},\ }\href
	{http://dx.doi.org/10.1038/srep38559} {\bibfield  {journal} {\bibinfo
			{journal} {Sci. Rep.}\ }\textbf {\bibinfo {volume} {6}},\ \bibinfo {pages}
		{38559} (\bibinfo {year} {2016})}\BibitemShut {NoStop}%
	\bibitem [{\citenamefont {Karvounis}\ \emph {et~al.}(2015)\citenamefont
		{Karvounis}, \citenamefont {Ou}, \citenamefont {Wu}, \citenamefont
		{MacDonald},\ and\ \citenamefont {Zheludev}}]{APL.107.191110}%
	\BibitemOpen
	\bibfield  {author} {\bibinfo {author} {\bibfnamefont {A.}~\bibnamefont
			{Karvounis}}, \bibinfo {author} {\bibfnamefont {J. Y.}\ \bibnamefont {Ou}},
		\bibinfo {author} {\bibfnamefont {W.}~\bibnamefont {Wu}}, \bibinfo {author}
		{\bibfnamefont {K.~F.}\ \bibnamefont {MacDonald}}, \ and\ \bibinfo {author}
		{\bibfnamefont {N.~I.}\ \bibnamefont {Zheludev}},\ }\href {\doibase
		10.1063/1.4935795} {\bibfield  {journal} {\bibinfo  {journal} {Appl. Phys.
				Lett.}\ }\textbf {\bibinfo {volume} {107}},\ \bibinfo {pages} {191110}
		(\bibinfo {year} {2015})}\BibitemShut {NoStop}%
	\bibitem [{\citenamefont {Chan}\ \emph {et~al.}(2011)\citenamefont {Chan},
		\citenamefont {Alegre}, \citenamefont {Safavi-Naeini}, \citenamefont {Hill},
		\citenamefont {Krause}, \citenamefont {Gr\"oblacher}, \citenamefont
		{Aspelmeyer},\ and\ \citenamefont {Painter}}]{Nature.478.89}%
	\BibitemOpen
	\bibfield  {author} {\bibinfo {author} {\bibfnamefont {J.}~\bibnamefont
			{Chan}}, \bibinfo {author} {\bibfnamefont {T.~P.~M.}\ \bibnamefont {Alegre}},
		\bibinfo {author} {\bibfnamefont {A.~H.}\ \bibnamefont {Safavi-Naeini}},
		\bibinfo {author} {\bibfnamefont {J.~T.}\ \bibnamefont {Hill}}, \bibinfo
		{author} {\bibfnamefont {A.}~\bibnamefont {Krause}}, \bibinfo {author}
		{\bibfnamefont {S.}~\bibnamefont {Gr\"oblacher}}, \bibinfo {author}
		{\bibfnamefont {M.}~\bibnamefont {Aspelmeyer}}, \ and\ \bibinfo {author}
		{\bibfnamefont {O.}~\bibnamefont {Painter}},\ }\href
	{http://dx.doi.org/10.1038/nature10461} {\bibfield  {journal} {\bibinfo
			{journal} {Nature (London)}\ }\textbf {\bibinfo {volume} {478}},\ \bibinfo
		{pages} {89} (\bibinfo {year} {2011})}\BibitemShut {NoStop}%
	\bibitem [{\citenamefont {Wollman}\ \emph {et~al.}(2015)\citenamefont
		{Wollman}, \citenamefont {Lei}, \citenamefont {Weinstein}, \citenamefont
		{Suh}, \citenamefont {Kronwald}, \citenamefont {Marquardt}, \citenamefont
		{Clerk},\ and\ \citenamefont {Schwab}}]{Science.349.952}%
	\BibitemOpen
	\bibfield  {author} {\bibinfo {author} {\bibfnamefont {E.~E.}\ \bibnamefont
			{Wollman}}, \bibinfo {author} {\bibfnamefont {C.~U.}\ \bibnamefont {Lei}},
		\bibinfo {author} {\bibfnamefont {A.~J.}\ \bibnamefont {Weinstein}}, \bibinfo
		{author} {\bibfnamefont {J.}~\bibnamefont {Suh}}, \bibinfo {author}
		{\bibfnamefont {A.}~\bibnamefont {Kronwald}}, \bibinfo {author}
		{\bibfnamefont {F.}~\bibnamefont {Marquardt}}, \bibinfo {author}
		{\bibfnamefont {A.~A.}\ \bibnamefont {Clerk}}, \ and\ \bibinfo {author}
		{\bibfnamefont {K.~C.}\ \bibnamefont {Schwab}},\ }\href {\doibase
		10.1126/science.aac5138} {\bibfield  {journal} {\bibinfo  {journal}
			{Science}\ }\textbf {\bibinfo {volume} {349}},\ \bibinfo {pages} {952}
		(\bibinfo {year} {2015})}\BibitemShut {NoStop}%
	\bibitem [{\citenamefont {Agarwal}\ and\ \citenamefont
		{Huang}(2010)}]{PhysRevA.81.041803}%
	\BibitemOpen
	\bibfield  {author} {\bibinfo {author} {\bibfnamefont {G.~S.}\ \bibnamefont
			{Agarwal}}\ and\ \bibinfo {author} {\bibfnamefont {S.}~\bibnamefont
			{Huang}},\ }\href {\doibase 10.1103/PhysRevA.81.041803} {\bibfield  {journal}
		{\bibinfo  {journal} {Phys. Rev. A}\ }\textbf {\bibinfo {volume} {81}},\
		\bibinfo {pages} {041803(R)} (\bibinfo {year} {2010})}\BibitemShut {NoStop}%
	\bibitem [{\citenamefont {Weis}\ \emph {et~al.}(2010)\citenamefont {Weis},
		\citenamefont {Rivi{\`e}re}, \citenamefont {Del{\'e}glise}, \citenamefont
		{Gavartin}, \citenamefont {Arcizet}, \citenamefont {Schliesser},\ and\
		\citenamefont {Kippenberg}}]{Science.330.1520}%
	\BibitemOpen
	\bibfield  {author} {\bibinfo {author} {\bibfnamefont {S.}~\bibnamefont
			{Weis}}, \bibinfo {author} {\bibfnamefont {R.}~\bibnamefont {Rivi{\`e}re}},
		\bibinfo {author} {\bibfnamefont {S.}~\bibnamefont {Del{\'e}glise}}, \bibinfo
		{author} {\bibfnamefont {E.}~\bibnamefont {Gavartin}}, \bibinfo {author}
		{\bibfnamefont {O.}~\bibnamefont {Arcizet}}, \bibinfo {author} {\bibfnamefont
			{A.}~\bibnamefont {Schliesser}}, \ and\ \bibinfo {author} {\bibfnamefont
			{T.~J.}\ \bibnamefont {Kippenberg}},\ }\href {\doibase
		10.1126/science.1195596} {\bibfield  {journal} {\bibinfo  {journal}
			{Science}\ }\textbf {\bibinfo {volume} {330}},\ \bibinfo {pages} {1520}
		(\bibinfo {year} {2010})}\BibitemShut {NoStop}%
	\bibitem [{\citenamefont {Rabl}(2011)}]{PhysRevLett.107.063601}%
	\BibitemOpen
	\bibfield  {author} {\bibinfo {author} {\bibfnamefont {P.}~\bibnamefont
			{Rabl}},\ }\href {\doibase 10.1103/PhysRevLett.107.063601} {\bibfield
		{journal} {\bibinfo  {journal} {Phys. Rev. Lett.}\ }\textbf {\bibinfo
			{volume} {107}},\ \bibinfo {pages} {063601} (\bibinfo {year}
		{2011})}\BibitemShut {NoStop}%
	\bibitem [{\citenamefont {Nunnenkamp}, \citenamefont {B\o{}rkje},\ and\
		\citenamefont {Girvin}(2011)}]{PhysRevLett.107.063602}%
	\BibitemOpen
	\bibfield  {author} {\bibinfo {author} {\bibfnamefont {A.}~\bibnamefont
			{Nunnenkamp}}, \bibinfo {author} {\bibfnamefont {K.}~\bibnamefont
			{B\o{}rkje}}, \ and\ \bibinfo {author} {\bibfnamefont {S.~M.}\ \bibnamefont
			{Girvin}},\ }\href {\doibase 10.1103/PhysRevLett.107.063602} {\bibfield
		{journal} {\bibinfo  {journal} {Phys. Rev. Lett.}\ }\textbf {\bibinfo
			{volume} {107}},\ \bibinfo {pages} {063602} (\bibinfo {year}
		{2011})}\BibitemShut {NoStop}%
	\bibitem [{\citenamefont {Liao}\ and\ \citenamefont
		{Nori}(2013)}]{PhysRevA.88.023853}%
	\BibitemOpen
	\bibfield  {author} {\bibinfo {author} {\bibfnamefont {J. Q.}\ \bibnamefont
			{Liao}}\ and\ \bibinfo {author} {\bibfnamefont {F.}~\bibnamefont {Nori}},\
	}\href {\doibase 10.1103/PhysRevA.88.023853} {\bibfield  {journal} {\bibinfo
			{journal} {Phys. Rev. A}\ }\textbf {\bibinfo {volume} {88}},\ \bibinfo
		{pages} {023853} (\bibinfo {year} {2013})}\BibitemShut {NoStop}%
	\bibitem [{\citenamefont {Xu}\ and\ \citenamefont {Li}(2013)}]{JPB.46.035502}%
	\BibitemOpen
	\bibfield  {author} {\bibinfo {author} {\bibfnamefont {X. W.}\ \bibnamefont
			{Xu}}\ and\ \bibinfo {author} {\bibfnamefont {Y. J.}\ \bibnamefont {Li}},\
	}\href {http://stacks.iop.org/0953-4075/46/i=3/a=035502} {\bibfield
		{journal} {\bibinfo  {journal} {J. Phys. B: At., Mol. Opt. Phys.}\ }\textbf
		{\bibinfo {volume} {46}},\ \bibinfo {pages} {035502} (\bibinfo {year}
		{2013})}\BibitemShut {NoStop}%
	\bibitem [{\citenamefont {Zhai}\ \emph {et~al.}(2019)\citenamefont {Zhai},
		\citenamefont {Huang}, \citenamefont {Jing},\ and\ \citenamefont
		{Kuang}}]{OE.27.27649}%
	\BibitemOpen
	\bibfield  {author} {\bibinfo {author} {\bibfnamefont {C.}~\bibnamefont
			{Zhai}}, \bibinfo {author} {\bibfnamefont {R.}~\bibnamefont {Huang}},
		\bibinfo {author} {\bibfnamefont {H.}~\bibnamefont {Jing}}, \ and\ \bibinfo
		{author} {\bibfnamefont {L. M.}\ \bibnamefont {Kuang}},\ }\href {\doibase
		10.1364/OE.27.027649} {\bibfield  {journal} {\bibinfo  {journal} {Opt.
				Express}\ }\textbf {\bibinfo {volume} {27}},\ \bibinfo {pages} {27649}
		(\bibinfo {year} {2019})}\BibitemShut {NoStop}%
	\bibitem [{\citenamefont {Wang}\ \emph {et~al.}(2019)\citenamefont {Wang},
		\citenamefont {Bai}, \citenamefont {Liu}, \citenamefont {Zhang},\ and\
		\citenamefont {Wang}}]{PhysRevA.99.043818}%
	\BibitemOpen
	\bibfield  {author} {\bibinfo {author} {\bibfnamefont {D. Y.}\ \bibnamefont
			{Wang}}, \bibinfo {author} {\bibfnamefont {C. H.}\ \bibnamefont {Bai}},
		\bibinfo {author} {\bibfnamefont {S.}~\bibnamefont {Liu}}, \bibinfo {author}
		{\bibfnamefont {S.}~\bibnamefont {Zhang}}, \ and\ \bibinfo {author}
		{\bibfnamefont {H. F.}\ \bibnamefont {Wang}},\ }\href {\doibase
		10.1103/PhysRevA.99.043818} {\bibfield  {journal} {\bibinfo  {journal} {Phys.
				Rev. A}\ }\textbf {\bibinfo {volume} {99}},\ \bibinfo {pages} {043818}
		(\bibinfo {year} {2019})}\BibitemShut {NoStop}%
	\bibitem [{\citenamefont {Miranowicz}\ \emph {et~al.}(2013)\citenamefont
		{Miranowicz}, \citenamefont {Paprzycka}, \citenamefont {Liu}, \citenamefont
		{Bajer},\ and\ \citenamefont {Nori}}]{PhysRevA.87.023809}%
	\BibitemOpen
	\bibfield  {author} {\bibinfo {author} {\bibfnamefont {A.}~\bibnamefont
			{Miranowicz}}, \bibinfo {author} {\bibfnamefont {M.}~\bibnamefont
			{Paprzycka}}, \bibinfo {author} {\bibfnamefont {Y. X.}\ \bibnamefont {Liu}},
		\bibinfo {author} {\bibfnamefont {J.}~\bibnamefont {Bajer}}, \ and\ \bibinfo
		{author} {\bibfnamefont {F.}~\bibnamefont {Nori}},\ }\href {\doibase
		10.1103/PhysRevA.87.023809} {\bibfield  {journal} {\bibinfo  {journal} {Phys.
				Rev. A}\ }\textbf {\bibinfo {volume} {87}},\ \bibinfo {pages} {023809}
		(\bibinfo {year} {2013})}\BibitemShut {NoStop}%
	\bibitem [{\citenamefont {Hamsen}\ \emph {et~al.}(2017)\citenamefont {Hamsen},
		\citenamefont {Tolazzi}, \citenamefont {Wilk},\ and\ \citenamefont
		{Rempe}}]{PhysRevLett.118.133604}%
	\BibitemOpen
	\bibfield  {author} {\bibinfo {author} {\bibfnamefont {C.}~\bibnamefont
			{Hamsen}}, \bibinfo {author} {\bibfnamefont {K.~N.}\ \bibnamefont {Tolazzi}},
		\bibinfo {author} {\bibfnamefont {T.}~\bibnamefont {Wilk}}, \ and\ \bibinfo
		{author} {\bibfnamefont {G.}~\bibnamefont {Rempe}},\ }\href {\doibase
		10.1103/PhysRevLett.118.133604} {\bibfield  {journal} {\bibinfo  {journal}
			{Phys. Rev. Lett.}\ }\textbf {\bibinfo {volume} {118}},\ \bibinfo {pages}
		{133604} (\bibinfo {year} {2017})}\BibitemShut {NoStop}%
	\bibitem [{\citenamefont {Kowalewska-Kud\l{}aszyk}\ \emph
		{et~al.}(2019)\citenamefont {Kowalewska-Kud\l{}aszyk}, \citenamefont {Abo},
		\citenamefont {Chimczak}, \citenamefont {Pe\ifmmode~\check{r}\else
			\v{r}\fi{}ina}, \citenamefont {Nori},\ and\ \citenamefont
		{Miranowicz}}]{PhysRevA.100.053857}%
	\BibitemOpen
	\bibfield  {author} {\bibinfo {author} {\bibfnamefont {A.}~\bibnamefont
			{Kowalewska-Kud\l{}aszyk}}, \bibinfo {author} {\bibfnamefont {S.~I.}\
			\bibnamefont {Abo}}, \bibinfo {author} {\bibfnamefont {G.}~\bibnamefont
			{Chimczak}}, \bibinfo {author} {\bibfnamefont {J.}~\bibnamefont
			{Pe\ifmmode~\check{r}\else \v{r}\fi{}ina}}, \bibinfo {author} {\bibfnamefont
			{F.}~\bibnamefont {Nori}}, \ and\ \bibinfo {author} {\bibfnamefont
			{A.}~\bibnamefont {Miranowicz}},\ }\href {\doibase
		10.1103/PhysRevA.100.053857} {\bibfield  {journal} {\bibinfo  {journal}
			{Phys. Rev. A}\ }\textbf {\bibinfo {volume} {100}},\ \bibinfo {pages}
		{053857} (\bibinfo {year} {2019})}\BibitemShut {NoStop}%
	\bibitem [{\citenamefont {Tian}\ and\ \citenamefont
		{Carmichael}(1992)}]{PhysRevA.46.R6801}%
	\BibitemOpen
	\bibfield  {author} {\bibinfo {author} {\bibfnamefont {L.}~\bibnamefont
			{Tian}}\ and\ \bibinfo {author} {\bibfnamefont {H.~J.}\ \bibnamefont
			{Carmichael}},\ }\href {\doibase 10.1103/PhysRevA.46.R6801} {\bibfield
		{journal} {\bibinfo  {journal} {Phys. Rev. A}\ }\textbf {\bibinfo {volume}
			{46}},\ \bibinfo {pages} {R6801} (\bibinfo {year} {1992})}\BibitemShut
	{NoStop}%
	\bibitem [{\citenamefont {Imamo\ifmmode~\bar{g}\else \={g}\fi{}lu}\ \emph
		{et~al.}(1997)\citenamefont {Imamo\ifmmode~\bar{g}\else \={g}\fi{}lu},
		\citenamefont {Schmidt}, \citenamefont {Woods},\ and\ \citenamefont
		{Deutsch}}]{PhysRevLett.79.1467}%
	\BibitemOpen
	\bibfield  {author} {\bibinfo {author} {\bibfnamefont {A.}~\bibnamefont
			{Imamo\ifmmode~\bar{g}\else \={g}\fi{}lu}}, \bibinfo {author} {\bibfnamefont
			{H.}~\bibnamefont {Schmidt}}, \bibinfo {author} {\bibfnamefont
			{G.}~\bibnamefont {Woods}}, \ and\ \bibinfo {author} {\bibfnamefont
			{M.}~\bibnamefont {Deutsch}},\ }\href {\doibase 10.1103/PhysRevLett.79.1467}
	{\bibfield  {journal} {\bibinfo  {journal} {Phys. Rev. Lett.}\ }\textbf
		{\bibinfo {volume} {79}},\ \bibinfo {pages} {1467} (\bibinfo {year}
		{1997})}\BibitemShut {NoStop}%
	\bibitem [{\citenamefont {Shen}, \citenamefont {Zhou},\ and\ \citenamefont
		{Yi}(2014)}]{PhysRevA.90.023849}%
	\BibitemOpen
	\bibfield  {author} {\bibinfo {author} {\bibfnamefont {H.~Z.}\ \bibnamefont
			{Shen}}, \bibinfo {author} {\bibfnamefont {Y.~H.}\ \bibnamefont {Zhou}}, \
		and\ \bibinfo {author} {\bibfnamefont {X.~X.}\ \bibnamefont {Yi}},\ }\href
	{\doibase 10.1103/PhysRevA.90.023849} {\bibfield  {journal} {\bibinfo
			{journal} {Phys. Rev. A}\ }\textbf {\bibinfo {volume} {90}},\ \bibinfo
		{pages} {023849} (\bibinfo {year} {2014})}\BibitemShut {NoStop}%
	\bibitem [{\citenamefont {Liew}\ and\ \citenamefont
		{Savona}(2010)}]{PhysRevLett.104.183601}%
	\BibitemOpen
	\bibfield  {author} {\bibinfo {author} {\bibfnamefont {T.~C.~H.}\
			\bibnamefont {Liew}}\ and\ \bibinfo {author} {\bibfnamefont {V.}~\bibnamefont
			{Savona}},\ }\href {\doibase 10.1103/PhysRevLett.104.183601} {\bibfield
		{journal} {\bibinfo  {journal} {Phys. Rev. Lett.}\ }\textbf {\bibinfo
			{volume} {104}},\ \bibinfo {pages} {183601} (\bibinfo {year}
		{2010})}\BibitemShut {NoStop}%
	\bibitem [{\citenamefont {Sarma}\ and\ \citenamefont
		{Sarma}(2017)}]{PhysRevA.96.053827}%
	\BibitemOpen
	\bibfield  {author} {\bibinfo {author} {\bibfnamefont {B.}~\bibnamefont
			{Sarma}}\ and\ \bibinfo {author} {\bibfnamefont {A.~K.}\ \bibnamefont
			{Sarma}},\ }\href {\doibase 10.1103/PhysRevA.96.053827} {\bibfield  {journal}
		{\bibinfo  {journal} {Phys. Rev. A}\ }\textbf {\bibinfo {volume} {96}},\
		\bibinfo {pages} {053827} (\bibinfo {year} {2017})}\BibitemShut {NoStop}%
	\bibitem [{\citenamefont {Trivedi}\ \emph {et~al.}(2019)\citenamefont
		{Trivedi}, \citenamefont {Radulaski}, \citenamefont {Fischer}, \citenamefont
		{Fan},\ and\ \citenamefont {Vu\ifmmode \check{c}\else
			\v{c}\fi{}kovi\ifmmode~\acute{c}\else \'{c}\fi{}}}]{PhysRevLett.122.243602}%
	\BibitemOpen
	\bibfield  {author} {\bibinfo {author} {\bibfnamefont {R.}~\bibnamefont
			{Trivedi}}, \bibinfo {author} {\bibfnamefont {M.}~\bibnamefont {Radulaski}},
		\bibinfo {author} {\bibfnamefont {K.~A.}\ \bibnamefont {Fischer}}, \bibinfo
		{author} {\bibfnamefont {S.}~\bibnamefont {Fan}}, \ and\ \bibinfo {author}
		{\bibfnamefont {J.}~\bibnamefont {Vu\ifmmode \check{c}\else
				\v{c}\fi{}kovi\ifmmode~\acute{c}\else \'{c}\fi{}}},\ }\href {\doibase
		10.1103/PhysRevLett.122.243602} {\bibfield  {journal} {\bibinfo  {journal}
			{Phys. Rev. Lett.}\ }\textbf {\bibinfo {volume} {122}},\ \bibinfo {pages}
		{243602} (\bibinfo {year} {2019})}\BibitemShut {NoStop}%
	\bibitem [{\citenamefont {Yan}\ \emph {et~al.}(2018)\citenamefont {Yan},
		\citenamefont {Cheng}, \citenamefont {Guan}, \citenamefont {Yu},\ and\
		\citenamefont {Duan}}]{OL.43.5086}%
	\BibitemOpen
	\bibfield  {author} {\bibinfo {author} {\bibfnamefont {Y.}~\bibnamefont
			{Yan}}, \bibinfo {author} {\bibfnamefont {Y.}~\bibnamefont {Cheng}}, \bibinfo
		{author} {\bibfnamefont {S.}~\bibnamefont {Guan}}, \bibinfo {author}
		{\bibfnamefont {D.}~\bibnamefont {Yu}}, \ and\ \bibinfo {author}
		{\bibfnamefont {Z.}~\bibnamefont {Duan}},\ }\href {\doibase
		10.1364/OL.43.005086} {\bibfield  {journal} {\bibinfo  {journal} {Opt.
				Lett.}\ }\textbf {\bibinfo {volume} {43}},\ \bibinfo {pages} {5086} (\bibinfo
		{year} {2018})}\BibitemShut {NoStop}%
	\bibitem [{\citenamefont {Flayac}\ and\ \citenamefont
		{Savona}(2017)}]{PhysRevA.96.053810}%
	\BibitemOpen
	\bibfield  {author} {\bibinfo {author} {\bibfnamefont {H.}~\bibnamefont
			{Flayac}}\ and\ \bibinfo {author} {\bibfnamefont {V.}~\bibnamefont
			{Savona}},\ }\href {\doibase 10.1103/PhysRevA.96.053810} {\bibfield
		{journal} {\bibinfo  {journal} {Phys. Rev. A}\ }\textbf {\bibinfo {volume}
			{96}},\ \bibinfo {pages} {053810} (\bibinfo {year} {2017})}\BibitemShut
	{NoStop}%
	\bibitem [{\citenamefont {K\'om\'ar}\ \emph {et~al.}(2013)\citenamefont
		{K\'om\'ar}, \citenamefont {Bennett}, \citenamefont {Stannigel},
		\citenamefont {Habraken}, \citenamefont {Rabl}, \citenamefont {Zoller},\ and\
		\citenamefont {Lukin}}]{PhysRevA.87.013839}%
	\BibitemOpen
	\bibfield  {author} {\bibinfo {author} {\bibfnamefont {P.}~\bibnamefont
			{K\'om\'ar}}, \bibinfo {author} {\bibfnamefont {S.~D.}\ \bibnamefont
			{Bennett}}, \bibinfo {author} {\bibfnamefont {K.}~\bibnamefont {Stannigel}},
		\bibinfo {author} {\bibfnamefont {S.~J.~M.}\ \bibnamefont {Habraken}},
		\bibinfo {author} {\bibfnamefont {P.}~\bibnamefont {Rabl}}, \bibinfo {author}
		{\bibfnamefont {P.}~\bibnamefont {Zoller}}, \ and\ \bibinfo {author}
		{\bibfnamefont {M.~D.}\ \bibnamefont {Lukin}},\ }\href {\doibase
		10.1103/PhysRevA.87.013839} {\bibfield  {journal} {\bibinfo  {journal} {Phys.
				Rev. A}\ }\textbf {\bibinfo {volume} {87}},\ \bibinfo {pages} {013839}
		(\bibinfo {year} {2013})}\BibitemShut {NoStop}%
	\bibitem [{\citenamefont {Boyd}(2003)}]{NonlinearOptics}%
	\BibitemOpen
	\bibfield  {author} {\bibinfo {author} {\bibfnamefont {R.~W.}\ \bibnamefont
			{Boyd}},\ }\href@noop {} {\emph {\bibinfo {title} {Nonlinear Optics}}}\
	(\bibinfo  {publisher} {Academic press},\ \bibinfo {year} {2003})\BibitemShut
	{NoStop}%
	\bibitem [{\citenamefont {Fu}\ \emph {et~al.}(2018)\citenamefont {Fu},
		\citenamefont {Xue}, \citenamefont {Midorikawa},\ and\ \citenamefont
		{Takahashi}}]{APL.112.241105}%
	\BibitemOpen
	\bibfield  {author} {\bibinfo {author} {\bibfnamefont {Y.}~\bibnamefont
			{Fu}}, \bibinfo {author} {\bibfnamefont {B.}~\bibnamefont {Xue}}, \bibinfo
		{author} {\bibfnamefont {K.}~\bibnamefont {Midorikawa}}, \ and\ \bibinfo
		{author} {\bibfnamefont {E.~J.}\ \bibnamefont {Takahashi}},\ }\href {\doibase
		10.1063/1.5038414} {\bibfield  {journal} {\bibinfo  {journal} {Appl. 
				Phys. Lett.}\ }\textbf {\bibinfo {volume} {112}},\ \bibinfo {pages}
		{241105} (\bibinfo {year} {2018})}\BibitemShut {NoStop}%
	\bibitem [{\citenamefont {Yin}\ \emph {et~al.}(2018)\citenamefont {Yin},
		\citenamefont {L\"{u}}, \citenamefont {Wan}, \citenamefont {Bin},\ and\
		\citenamefont {Wu}}]{OL.43.2050}%
	\BibitemOpen
	\bibfield  {author} {\bibinfo {author} {\bibfnamefont {T. S.}\ \bibnamefont
			{Yin}}, \bibinfo {author} {\bibfnamefont {X. Y.}\ \bibnamefont {L\"{u}}},
		\bibinfo {author} {\bibfnamefont {L. L.}\ \bibnamefont {Wan}}, \bibinfo
		{author} {\bibfnamefont {S. W.}\ \bibnamefont {Bin}}, \ and\ \bibinfo
		{author} {\bibfnamefont {Y.}~\bibnamefont {Wu}},\ }\href {\doibase
		10.1364/OL.43.002050} {\bibfield  {journal} {\bibinfo  {journal} {Opt.
				Lett.}\ }\textbf {\bibinfo {volume} {43}},\ \bibinfo {pages} {2050} (\bibinfo
		{year} {2018})}\BibitemShut {NoStop}%
	\bibitem [{\citenamefont {Agarwal}\ and\ \citenamefont
		{Huang}(2016)}]{PhysRevA.93.043844}%
	\BibitemOpen
	\bibfield  {author} {\bibinfo {author} {\bibfnamefont {G.~S.}\ \bibnamefont
			{Agarwal}}\ and\ \bibinfo {author} {\bibfnamefont {S.}~\bibnamefont
			{Huang}},\ }\href {\doibase 10.1103/PhysRevA.93.043844} {\bibfield  {journal}
		{\bibinfo  {journal} {Phys. Rev. A}\ }\textbf {\bibinfo {volume} {93}},\
		\bibinfo {pages} {043844} (\bibinfo {year} {2016})}\BibitemShut {NoStop}%
	\bibitem [{\citenamefont {Poot}\ and\ \citenamefont
		{Tang}(2014)}]{APL.104.061101}%
	\BibitemOpen
	\bibfield  {author} {\bibinfo {author} {\bibfnamefont {M.}~\bibnamefont
			{Poot}}\ and\ \bibinfo {author} {\bibfnamefont {H.~X.}\ \bibnamefont
			{Tang}},\ }\href {\doibase 10.1063/1.4864257} {\bibfield  {journal} {\bibinfo
			{journal} {Appl. Phys. Lett.}\ }\textbf {\bibinfo {volume} {104}},\ \bibinfo
		{pages} {061101} (\bibinfo {year} {2014})}\BibitemShut {NoStop}%
	\bibitem [{\citenamefont {Wang}\ \emph {et~al.}(2018)\citenamefont {Wang},
		\citenamefont {Bai}, \citenamefont {Liu}, \citenamefont {Zhang},\ and\
		\citenamefont {Wang}}]{PhysRevA.98.023816}%
	\BibitemOpen
	\bibfield  {author} {\bibinfo {author} {\bibfnamefont {D. Y.}\ \bibnamefont
			{Wang}}, \bibinfo {author} {\bibfnamefont {C. H.}\ \bibnamefont {Bai}},
		\bibinfo {author} {\bibfnamefont {S.}~\bibnamefont {Liu}}, \bibinfo {author}
		{\bibfnamefont {S.}~\bibnamefont {Zhang}}, \ and\ \bibinfo {author}
		{\bibfnamefont {H. F.}\ \bibnamefont {Wang}},\ }\href {\doibase
		10.1103/PhysRevA.98.023816} {\bibfield  {journal} {\bibinfo  {journal} {Phys.
				Rev. A}\ }\textbf {\bibinfo {volume} {98}},\ \bibinfo {pages} {023816}
		(\bibinfo {year} {2018})}\BibitemShut {NoStop}%
	\bibitem [{\citenamefont {Xiong}\ \emph {et~al.}(2018)\citenamefont {Xiong},
		\citenamefont {Li}, \citenamefont {Chao},\ and\ \citenamefont
		{Zhou}}]{OL43.6053}%
	\BibitemOpen
	\bibfield  {author} {\bibinfo {author} {\bibfnamefont {B.}~\bibnamefont
			{Xiong}}, \bibinfo {author} {\bibfnamefont {X.}~\bibnamefont {Li}}, \bibinfo
		{author} {\bibfnamefont {S. L.}\ \bibnamefont {Chao}}, \ and\ \bibinfo
		{author} {\bibfnamefont {L.}~\bibnamefont {Zhou}},\ }\href {\doibase
		10.1364/OL.43.006053} {\bibfield  {journal} {\bibinfo  {journal} {Opt.
				Lett.}\ }\textbf {\bibinfo {volume} {43}},\ \bibinfo {pages} {6053} (\bibinfo
		{year} {2018})}\BibitemShut {NoStop}%
	\bibitem [{\citenamefont {Bai}\ \emph {et~al.}(2019)\citenamefont {Bai},
		\citenamefont {Wang}, \citenamefont {Zhang}, \citenamefont {Liu},\ and\
		\citenamefont {Wang}}]{PRJ.7.1229}%
	\BibitemOpen
	\bibfield  {author} {\bibinfo {author} {\bibfnamefont {C. H.}\ \bibnamefont
			{Bai}}, \bibinfo {author} {\bibfnamefont {D. Y.}\ \bibnamefont {Wang}},
		\bibinfo {author} {\bibfnamefont {S.}~\bibnamefont {Zhang}}, \bibinfo
		{author} {\bibfnamefont {S.}~\bibnamefont {Liu}}, \ and\ \bibinfo {author}
		{\bibfnamefont {H. F.}\ \bibnamefont {Wang}},\ }\href {\doibase
		10.1364/PRJ.7.001229} {\bibfield  {journal} {\bibinfo  {journal} {Photon.
				Res.}\ }\textbf {\bibinfo {volume} {7}},\ \bibinfo {pages} {1229} (\bibinfo
		{year} {2019})}\BibitemShut {NoStop}%
\end{thebibliography}
%
\end{document}